\documentclass[11pt]{article}
\usepackage{hyperref}
\pdfoutput=1
\begin{document}
\title{Feeling, following, feeding, fleeing: a copepod's life at low Reynolds number}
\author{David Murphy, Rachel Lasley, Donald Webster, Jeannette Yen \\
\\\vspace{6pt} \\ Georgia Institute of Technology, Atlanta, GA 30332, USA}
\maketitle
\begin{abstract}
In this fluid dynamics video, we present various aspects of copepod behavior at low\textit{ Re}.
\end{abstract}
\section{Introduction}

Copepods are small free-swimming crustaceans that are well adapted to life at a Reynolds number of $1$ to $ 1000$. Many aspects of their behavior, from swimming to feeding and escaping to mating, are influenced by the physics of the flow in this viscous regime and by their ability to sense that fluid environment. Copepods feel their surroundings through their long antennules that bear fine setae that function as flow sensors. Flow disturbances by prey, predators, or mates can be decoded by the copepod to inform them of their environment, and strain rates as low as $0.05\; \hbox{s}^{-1}$ can be detected.  In this work, we present high speed tomographic PIV videos (taken at $500$ fps) that demonstrate various aspects of the surprisingly sophisticated fluid-influenced behavior in one freshwater copepod, \textit{Hesperodiaptomus shoshone}, collected from alpine lakes high in the Rocky Mountains of Montana.\\
 
\textit{Hesperodiaptomus shoshone} are predatory copepods that feed on smaller copepods and on daphnids (water fleas). Two PIV video sequences demonstrate their ability to detect and attack their prey. In both cases, the flow disturbance generated by the moving prey initiates the attack by\textit{ H. shoshone}. In the first case, the antennal beat of the water flea (\textit{Daphnia middendorffiana}) clues the copepod in to its location. In the second video, the smaller copepod (\textit{H. kenai}) senses the approach of the larger \textit{H. shoshone} and initiates an escape jump. The larger copepod then attempts to capture its prey with its cephalic appendages. In both cases, \textit{H. shoshone} is unable to grasp its prey, and the prey escape.\\

Copepods, however, are not at the top of the food chain, and\textit{ H. shoshone} also serve as prey for fish. In order to study how copepods detect and flee from fish predators, we built a fish mimic. This impulsive siphon mimics the aquatic suction feeding technique that fish use to attack their prey. The second PIV sequence demonstrates copepods' sensitivity to minute flow signals and their ability to perform high speed escapes. The tomographic PIV system can be used to measure the flow structure surrounding the antennules in the milliseconds preceding the animal's escape in order to define a strain rate or vorticity escape threshold. When performing an escape jump with a speed of up to $1$ m/s, a copepod leaps into a regime where the effects of viscosity are much less important, and the Reynolds number may reach up to $1000 $. While most copepods successfully escaped, several animals were not so lucky and were captured by the fish mimic. These copepods may have been oriented such that they failed to perceive the fluid stimulus or they may have escaped in the "wrong" direction.\\

Understanding the behavior of zooplankton such as copepods requires a fundamental understanding of the fluid mechanics involved in low Reynolds number flows, an understanding enhanced by tools such as high speed tomographic PIV.

\end{document}